Scientific Research

# Analytical Evaluation of Space-Time M-QAM Signaling and Time-Delay Estimation in Multipath Terrestrial Communications

**Ardavan Rahimian**

School of Electronic, Electrical and Computer Engineering, University of Birmingham, Birmingham, UK.
Email: rahimian@ieee.org

## ABSTRACT

This paper presents novel information-theoretic PHY layer performance improvement for combating the system fading effects in wireless terrestrial communication applications. Robust space-time system strategies in cooperation with the high performance unpunctured serially concatenated codes have been jointly employed. The wireless terrestrial system also uses M-ary quadrature amplitude modulation (QAM) signaling to develop a well-established system modeling for multiple-input single-output (MISO) and multiple-input multiple-output (MIMO) analysis. This research also undertakes the analytical evaluation of time delay estimation in the presence of additive white Gaussian noise (AWGN) in multipath areas. The objective of this investigation is to develop mathematically-oriented estimations intended for a suitable employment in terrestrial communications. This model considers the problem of overlapping the terrestrial signals arrival-times estimation from a noisy received waveform in order to develop a system model for terrestrial TDE.

## 1. Introduction

The increasing demands for improving physical (PHY) layer system performance of wireless systems constitute a thriving global industry. Hence, in order to fulfill PHY layer QoS requirements in modern terrestrial systems, the appropriate combination of wireless technologies for constituting an engineering trade-off in terms of the overall system's bit error rate (BER), and the estimated complexities should be considered. After the introduction of the Alamouti dual-antenna space-time codes [1] which appeal in terms of its performance and complexity, the proposed scheme motivated Tarokh to generalize the Alamouti's scheme to an arbitrary number of antennas [2, 3], leading to the concept of the orthogonal space-time block codes (STBCs) [4]. STBCs have then evolved at an exceptional rate and reached a state of maturity in theory after their breakthrough discovery; they quickly found their way into the practical applications and standards for enabling outstanding reliable technologies such as: LTE, WLAN, WMN, WiMAX, and RFID [5-8]. In order to satisfy the reliability requirements in the systems, high-performance forward error correction (FEC) codes are required in addition to transmit/receive spatial diversity. Turbo codes [9] brought about a wealth of theoretical knowledge and practical solutions for achieving coding gains with feasible decoding complexities. They have been adopted in various wireless standards due to near-capacity performance, including the ITU-proposed 3G standards [10], ETSI-DVB system [11], UMTS [12], and LTE [13]. In this research, performance analysis of the dual antenna space-time codes based on the Alamouti along with the Tarokh's orthogonal STBCs has been evaluated in cooperation with the high-performance turbo coding scheme. The performance evaluations have been carried out over multiple-input single-output (MISO) and multiple-input multiple-output (MIMO) systems. These systems are operating in the statistical wireless terrestrial communications propagation environment. Hence, terrestrial reliable wireless system deployment is necessary in order to provide highly accurate and efficient services, and is also of crucial importance for modern wireless technologies. The enormous growth in terrestrial communications and with the increasing demand for the high-performance services raise the need for systems analysis and planning for the intended environments to evaluate performance and its operation estimation. The problem of wireless time-delay estimation (TDE) is of considerable research interest in the wireless communications, which is related to the multipath in specified environments. Multipath causes a single transmitted signal originating from a source to arrive at a receiver via two or more different paths along with any combination of several distinct unequal length paths. These extra paths at a receiver can be synchronized with the mean scaling of the most dominant path and all the others are treated as the noise; then possible equalization strategies can be employed to improve the receiver system performance [14,15]. It is assumed that the transmitted radio signal and the number of paths in the multipath environment are known. The received radio signal is not just the transmitted one, but several delayed and amplitude-weighted versions of the transmitted signal [16].

## 2. Terrestrial Communication TDE Theory

The problem of multipath time-delay estimation (MTDE) in this research has been characterized using (1) as:

$$\underline{yields} \; r(t) = \sum_{1}^{M} a_k s(t - \tau_k) + \tilde{\omega}(t); \; 0 \le t \le T. \quad (1)$$

where $s(t)$ is the transmitted radio signal (system pulse); $a_k$ is the amplitude (attenuation value) for path $k$; $t_k$ is the time-delay for path $k$; $M$ is the number of different paths; and $\tilde{\omega}(t)$ is the white Gaussian noise (AWGN) in the multipath environment. The correlation receiver is used to obtain the time-delay estimates. If the TDEs are estimated in a multipath environment where paths are spaced closer than a correlation time of a transmitted radio signal, the receiver cannot separate the closely spaced delays. When the terrestrial radio structure is utilized for the estimation, fading effects can be alleviated and significant wireless system performance improvement can be achieved [17].

## 3. Terrestrial Communication System Model

**Figure 1** shows block diagram of the STBC coded wireless MIMO system of interest, employing QAM as the modulation scheme, and turbo codes as a state-of-the-art FEC. The information bit sequence is first fed into a turbo encoder and then modulated in a multilevel QAM modulator where the coded system bit stream is first blocked into a symbol of $\text{Log}_2^M$ data bit $i.e.$ four data bits in this case; it should be considered as an appropriate operating point in system power versus spectral efficiency trade-off, since 16-QAM have moderate theoretical bandwidth efficiency in M-QAM signaling scheme [4 (bit/sec)/Hz] that is 1 (bit/sec)/Hz for 2-QAM ($i.e.$ BPSK), and 8 (bit/sec)/Hz for 256-QAM, while yields moderate power efficiency compared with other M-QAM schemes according to the theoretical BER bounds for uncoded M-QAM over the additive white Gaussian noise (AWGN) channel, for resulting in the BER of $10^{-6}$, and 16-QAM operating SNR is 14.5 dB, while that is 10.5 dB and 23.5 dB for $M = 2$ and $M = 256$ schemes respectively. Hence, it has been introduced in various standards such as DVB-Terrestrial, HSDPA protocol, and IEEE 802.11a PHY for modulating OFDM subcarriers. The modulated symbols are mapped using space-time block encoder which has advantages of both the spatial diversity provided by multiple antennas, and the temporal diversity available with time-varying fading [18], without channel state information (CSI) [19]. At the receiver, the combiner combines received signals which are then sent to the maximum likelihood (ML) detector. The detected symbols are then demodulated by the multilevel QAM demodulator and are sent to the iterative turbo decoder for recovering the transmitted information bits in the terrestrial radio system transmissions.

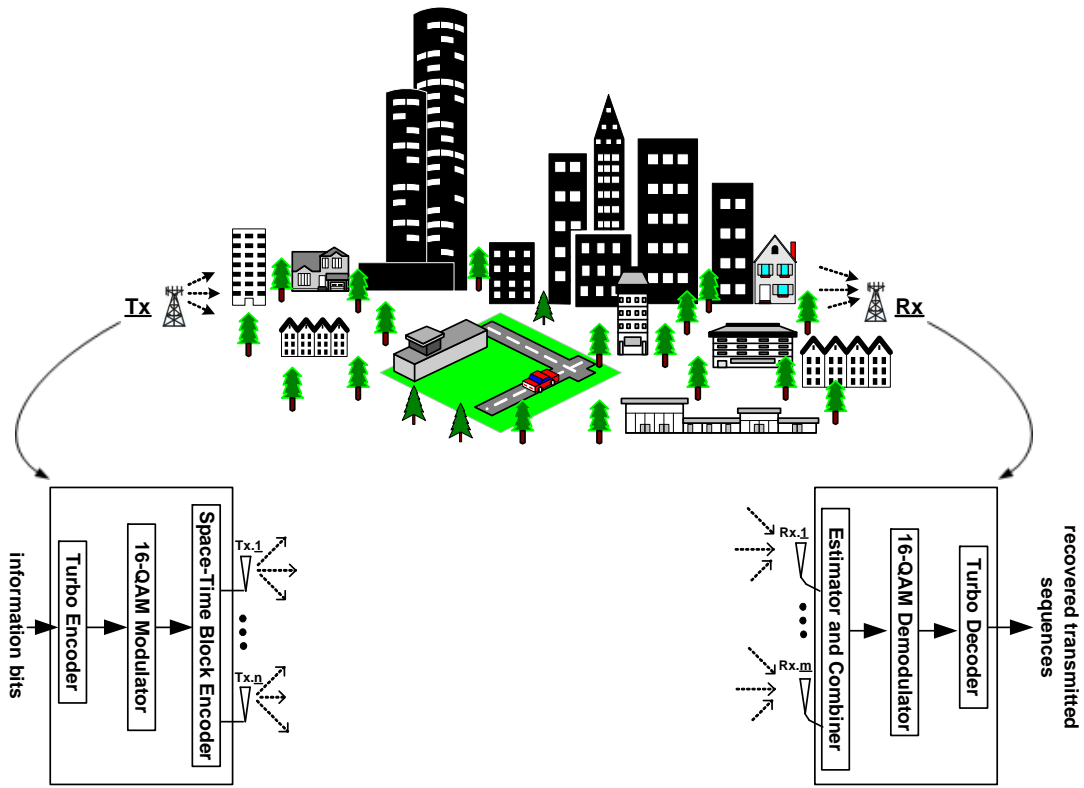

**Figure 1. Terrestrial communications environment block diagram.**

## 4. Terrestrial Communication FEC Coding

Turbo codes have attracted a great interest after their advent [9] since they were the first class of error correcting codes that could perform within a fraction of a decibel to capacity limit [20]. The classic turbo coding is a channel coding technique based on the idea of producing long and random code which employs an iterative decoding algorithm to result in BER performance near capacity limit. The encoder of classic turbo codes is constructed by parallel concatenating two recursive systematic convolutional (RSC) codes through an interleaver. They have been termed as parallel concatenated codes (PCCs) as well. The decoder consists of an interleaver linked soft-input soft-output constituent decoders for corresponding RSC codes. Owing to the presence of the system's interleaver, optimum ML decoding algorithms for turbo codes are very complex. However, iterative decoding of codes in which soft-information is exchanged between constituent decoders is quite feasible and leads to near capacity performance. This information could be regarded as a type of diversity that can refine the constituent decoder's outputs in iterations and in the TDE analysis [21]. An enhanced iterative turbo decoder structure using a modified version of the conventional maximum a posteriori (MAP) algorithm [22] was invoked to decode the introduced PCCs. For preserving the performance of the original MAP decoding with the relatively lower complexities, practical reduced complexity decoding approaches were then proposed. These system decoding algorithms introduced include Max Log-MAP [23], Log-MAP [24], and Erfanian scheme [25]. After achieving the system error correction characteristics close to the Shannon limit at a BER of $10^{-5}$ (at $E_b/N_0 = 0.7$ dB with code rate of 1/2) based on PCCs, a vast body of efforts are dedicated to develop the codes.

Unfortunately, in many turbo coded wireless systems, it is reported that for BERs lower than $10^{-5}$, the increase in transmit power just marginally reduces the BER values. As a strong technique to sharpen the system BER curve in error-floor region, turbo codes based on serial concatenation of convolutional codes were introduced which reported excellent results from this scheme in moderate to high SNRs [26]. Serial turbo codes which are referred to as serially concatenated codes (SCCs) behave significantly better and do not exhibit the typical error-floor behavior of the PCCs. In cases where relatively smaller interleaver size is employed, PCC's error-floor starts at a much higher BER value of $10^{-3}$, while the SCCs have shown to exhibit sharp curve for BERs lower than $10^{-3}$ [27]. The SCC's encoder consists of the cascade of an outer convolutional code, an interleaver, and also an inner convolutional code whose system input words are the permuted version of outer codewords. The turbo system decoding algorithm in both the SCCs and PCCs are the same. However, it should be noted that the primary difference is that while in the structure of PCCs both the constituent decoders make use of direct output from the demodulator, in SCC's decoder only the inner decoder makes use of this information directly, and the outer one receives the relative extrinsic system information from the deinterleaved codewords of inner decoder's outputs. In the context of designing SCCs, the following parameters have key roles in the turbo coded BER performance: constituent code; interleaver design; and decoding algorithm [28]. A practical performance parameter combination in the design of constituent codes and iterative decoder is utilized for yielding promising trade-off in terms of BER and complexity. SCCs have been also employed based on quadratic interleavers proposed in [29] which have not only shown to have better BER performance in both waterfall and error-floor regions compared with the classic interleavers [30], but also have straightforward system implementation [31]. In [30], extra coding gain of 2.5 - 3 dB for yielding the BER of $10^{-5}$ has been reported, when quadratic interleaver is used in SCCs for both the SISO and MISO systems over Rice multipath fading channel; compared with the wireless communication system using matrix-based block interleavers with large and comparable system depth and span.

## 5. Terrestrial Space-Time Coding

For suppressing multipath effects, diversity techniques like time, frequency, and spatial diversity could be employed. Recent multiple antenna systems based on the space-time codes in which devices are equipped with the multi-transmit and receive antennas, have received considerable attention for research, not only from the theoretical aspects, but also in development and implementation for wireless systems. They are expected to play a major role in futuristic wireless communication systems due to their high diversity order and low decoding complexities, which lead to significantly improve reliability and dramatically increase capacity of wireless systems links [32-35]. Space-time encoding is performed according to the number of the transmit antennas, and the code rate. For rate one of the twin-antenna space-time codes, denoting the signal transmitted from the first and second antennas by $s_1$ and $s_2$, the encoding and transmission sequences are shown in **Table 1**. The encoding and also transmission sequences for the half-rate three transmit antennas are presented in **Table 2**. The transmission model is based on transmitting signals $C_t^i, i = 1, 2, \ldots, n$ simultaneously from the $n$ transmit antennas at each time slot of $t$, under the assumption that the wireless channel is quasi-static (so that the path gains are considered to be constant over a frame of length $l$ and vary from one frame to the other one), and flat fading [3]. Defining the path gain from transmit antenna $i$ to the receive antenna $j$ by $\alpha_{i,j}$ and also denoting the noise samples by $\gamma_t^j$ (*i.e.* independent samples of a zero mean complex Gaussian random variable with variance $n/(2SNR)$ per complex system dimension), the received radio signal at the system antenna $j$ at the time $t$ is given as Equation (2) [3]:

**Table 1. Terrestrial Encoding and Transmission Sequence for Two Transmit Antennas; Alamouti Scheme; Code Rate: 1.**

| Time Slot | Antenna-I | Antenna-II |
|-----------|-----------|------------|
| I | $x_1$ | $x_2$ |
| II | $-x_2^*$ | $x_1^*$ |

**Table 2. Terrestrial Encoding and Transmission Sequence for Three Transmit Antennas; Tarokh Scheme; Code Rate: 1/2.**

| Time Slot | Antenna-I | Antenna-II | Antenna-III |
|-----------|-----------|------------|-------------|
| I | $x_1$ | $x_2$ | $x_3$ |
| II | $-x_2$ | $x_1$ | $-x_4$ |
| III | $-x_3$ | $x_4$ | $x_1$ |
| IV | $-x_4$ | $-x_3$ | $x_2$ |
| V | $x_1^*$ | $x_2^*$ | $x_3^*$ |
| VI | $-x_2^*$ | $x_1^*$ | $-x_4^*$ |
| VII | $-x_3^*$ | $x_4^*$ | $x_1^*$ |
| VIII | $-x_4^*$ | $-x_3^*$ | $x_2^*$ |

$$\text{yields } r_t^j = \sum_{i=1}^{n} \alpha_{i,j} c_t^i + \gamma_t^j. \tag{2}$$

The system's average energy of each transmitter antenna's symbols is normalized to one so that the average power of each receive antenna's radio signal is $n$. The receiver computes the decision metric given by the (3) over all the codewords, under the assumption that prefect channel impulse response (CSI) is available, and decides in favor of codeword that minimizes the sum [3]. The received signal is processed first in the channel estimator and combiner, and the combined received radio signals are then sent out to the system ML detector. In order for the wireless system to detect symbols of the two transmit antennas; denoted by $s_1$ and $s_2$; the decision metrics [36] have been potentially employed. For detecting symbols of half-rate three transmit antennas; denoted by $s_1$, $s_2$, $s_3$; the decoder minimizes the decision metrics, as in [36].

$$\text{yields } \sum_{t=1}^{l} \sum_{j=1}^{m} \left| r_t^j - \sum_{i=1}^{n} \alpha_{i,j} c_t^i \right|^2. \tag{3}$$

## 6. Terrestrial MIMO Channel Modeling

From system level aspect MIMO channel is represented by a $N \times M$ matrix $H$; the received terrestrial radio signal $y = [y_1, y_2, \ldots, y_m]^T$ can also be expressed as [37]: $y = H \cdot x + \hat{n} = \sum_{m=1}^{M} h_m \cdot x_m + \hat{n}$   $x = [x_1, x_2, \ldots, x_m]^T$ is the transmitted vector, $\hat{n}$ is the $N \times 1$ additive white Gaussian noise (AWGN). If the number of receive antennas is reduced to one (*i.e.* for MISO systems), the MISO's CIS can then be considered as $N$ SISO channels expressed as $1 \times N$ vector. It is assumed that $n_t$ transmit and $n_r$ receive antennas are placed in uniform linear arrays of normalized lengths $L_t$ and $L_r$ respectively. The normalized separation between transmit antennas $\Delta_t$ and the normalized separation between the receive antennas $\Delta_r$ are given: $\Delta_t = L_t / n_t$ and $\Delta_r = L_r / n_r$ respectively; where the system normalization is by the wavelength of the passband transmitted radio signal $\lambda_c$ [37]. It is then supposed there is an arbitrary number of physical paths between the transmitter and receiver, and the attenuation of path $\aleph$, and its angles with the transmit and receive system array are denoted as: $a_\aleph$, $\varphi_{t\aleph} (\omega_{t\aleph} := \cos \varphi_{t\aleph})$, and $\varphi_{r\aleph} (\omega_{r\aleph} := \cos \varphi_{r\aleph})$. Channel matrix $H$ is given as the following equations; where $d^{(i)}$ denotes the distance between the first transmit antenna 1 and the receive antenna 1, along the path $\aleph$ in radio transmission [37].

$$\text{yields } H = \sum_\aleph a_\aleph^b e_r (\omega_{r\aleph}) e_t (\omega_{t\aleph})^*; \tag{4}$$

$$\text{where } a_\aleph^b := a_\aleph \sqrt{n_t n_r} \exp\left(-\frac{j2\pi d^{(i)}}{\lambda_c}\right). \tag{5}$$

$$e_r(\omega) := \frac{1}{\sqrt{n_r}} \begin{bmatrix} 1 \\ \exp(-j2\pi\Delta_r\omega) \\ \vdots \\ \exp(-j2\pi(n_r-1)\Delta_r\omega) \end{bmatrix}. \quad (6)$$

$$e_t(\omega) := \frac{1}{\sqrt{n_t}} \begin{bmatrix} 1 \\ \exp(-j2\pi\Delta_t\omega) \\ \vdots \\ \exp(-j2\pi(n_t-1)\Delta_t\omega) \end{bmatrix}. \quad (7)$$

The column vectors: $e_t(\omega)$ and $e_r(\omega)$ are the transmitted and received unit spatial signatures along direction $\omega$ respectively. There are two primary applications for the channel models: location-specific models which provide very accurate models robust with respect to small errors, for optimizing a system in a certain geographical region (for the planning and deployment). Simplified channel models that describe the CIR statistics in a parametric form; in which the number of parameters is small, and is independent of the specific locations. The models which are reflecting the important properties of propagation channels are used for the design, testing, and type approval of wireless systems [38]. For comparing different MIMO systems and their relevant PHY layer performance enhancement techniques in a manner that is unified and agreed on by many system parties, various organizations introduce the reference channel models with reproducible conditions [38]. Among these modeling methods utilized in site-specific and generic channel model applications (*i.e.* stored CIR, deterministic channel models, and also stochastic system channel models), the latter one is then selected which describes the predicted probability density function (PDF) of the CIR over a large area. In this research, stochastic modeling of the received radio signal's amplitude is described by Rayleigh distribution which is used model for predicting PDF of the CIR; in the urban areas with blocked line-of-sight (LoS). Meanwhile, the COST207 models proposed bivariate Gaussian Doppler spectrum for modeling long echoes in urban propagation environments (models in the urban areas) [38]; given by:

$$P_G(f)$$
$$= \left(1/\sum_{i=1}^2 A_i\right) \left[\sum_{i=1}^2 \left(A_i/\sqrt{2\pi\sigma_i^2}\right) \cdot \exp\left(-(f-f_i)^2/2\sigma_i^2\right)\right];$$

where $f_i$, $A_i$, and $\sigma_i$, are center frequencies, power gains, and standard deviations respectively [35,39,40].

## 7. Terrestrial TDE Modeling and Analysis

In the case of complex and real amplitude error functions and in practice, the samples of the received and transmitted signals are existed for the time interval $0 < t < T$. The received terrestrial waveform can be modeled as (8):

$$r(nT_s) = \sum_{k=1}^M a_k s(nT_s - \tau_k) + \omega[n]; \ 0 \le n \le N-1. \quad (8)$$

where the sampling interval $T_s = T/N$ and $\omega[n]s$ are the noise samples of the AWGN with the terrestrial wireless system continuous autocorrelation as (9):

$$\underline{yields} \ R_{\tilde{\omega}}(\tau) = \sigma^2_{\tilde{\omega}}\delta(\tau). \quad (9)$$

where it is assumed that $\tilde{\omega}(t)$ is low-pass filtered (LPF) by the factor as in the Equation (10) below, in order to yield $\tilde{\omega}(t)$ such that the discrete system radio signal has a flat power spectrum as (11).

$$h_{L-P}(t) = \frac{1}{\pi t}\sin\left(\frac{\pi t}{T_s}\right) \overset{\tilde{\omega}(t)}{\Rightarrow} \omega[n] = \tilde{\omega}(nT_s); \quad (10)$$

$$\underline{yields} \ \sigma^2_{\tilde{\omega}} \equiv E\{\omega^2[n]\} = \frac{\sigma^2_{\tilde{\omega}}}{T_s}. \quad (11)$$

The least squares estimator (LSE) of $a = [a_1, a_2, \cdots, a_M]^T$ and $\tau = [\tau_1, \tau_2, \cdots, \tau_M]^T$ has then been obtained by minimizing the squared error function (12) as part of the system time-delay estimation (TDE) computation.

$$E_r(\tau,a) = \sum_{n\in N}\left|R[n] - S[n]\sum_{k=1}^M a_k e^{\frac{-j2\pi n\tau_k}{NT_s}}\right|^2; \quad (12)$$

$$\underline{yields} \ \|\tilde{r} - \tilde{p}(\lambda)a^2\|.$$

$$N = \left\{0 \le n \le \frac{N}{2} - 1; S[n] > threshold\right\} = \{q_1, \cdots, q_L\}. \ (13)$$

where $R[n]$ and $S[n]$ are the discrete Fourier transforms (DFTs) of $r[nT_s]$ and $s[nT_s]$, respectively. In the presence of AWGN, the LSE is also the maximum likelihood estimator (MLE). In the minimization procedure, a threshold is specified in order to employ only the spectrum regions with the satisfactory system SNRs. The weighting function is employed in order to enhance the accuracy of the estimated delay by attenuating the radiowaves fed into the correlator in spectral region where the SNR is the lowest. For the uncorrelated Gaussian radio system signal, noise, and single path, TDE is asymptotically unbiased and efficient in terms of the limit of long observation intervals. The terrestrial weights the cross-spectral phase according to the estimated phase when variance of the estimated phase error is the least. The system TDE procedure adapts the filter in order to insert a delay equal and opposite to that existing between the two radio sig-nals. In the case of the AWGN presence, the filter weight corresponding to true-delay obtains its maximum value compared to the other TDE terrestrial wireless system filter weights.

$$\lambda = \left[\lambda_1, \lambda_2, \cdots, \lambda_M\right]^T; \ \lambda_k = \frac{-2\pi\tau_k}{NT_s} \ \underline{yields}$$

$$r = \left[R[q_1], R[q_2], \cdots, R[q_L]\right]^T; \quad (14)$$

$$S = diag\left(S[q_1], \cdots, S[q_L]\right).$$

$$A(\lambda) = \begin{bmatrix} e^{j\lambda_1 q_1} & e^{j\lambda_2 q_1} & & e^{j\lambda_{M-1} q_1} & e^{j\lambda_M q_1} \\ e^{j\lambda_1 q_2} & e^{j\lambda_2 q_2} & \cdots & e^{j\lambda_{M-1} q_2} & e^{j\lambda_M q_2} \\ \vdots & & \ddots & & \vdots \\ e^{j\lambda_1 q_L} & e^{j\lambda_2 q_L} & \cdots & e^{j\lambda_{M-1} q_L} & e^{j\lambda_M q_L} \end{bmatrix}; \quad (15)$$

$$\underline{yields} \ \tilde{p}(\lambda) = SA(\lambda); \ \tilde{p}(\lambda) = \left[p(\lambda_1), \cdots, p(\lambda_M)\right].$$

where in the relation (14), $S[q_k]$ and $R[q_k]$ are the time-delay samples that lie above the specified terrestrial wireless system time-delay estimation threshold.

$$\tilde{P}(\lambda) = Q(\lambda)R(\lambda) \overset{QR \ Decomposition}{\Longrightarrow}$$
$$Q(\lambda) = \left[q_1(\lambda_1), q_2(\lambda_1, \lambda_2), \cdots, q_M(\lambda_1, \cdots, \lambda_M)\right]. \quad (16)$$

where in the relation (16), the $q_i$s are orthogonal, and each one is a function of $(\lambda_1, \lambda_2, \cdots, \lambda_i)$. Hence $E_r(\lambda)$ can be written as Equation (17) below, in order to minimize the error function for the terrestrial radio signal time-delays.

$$E_r(\lambda) = \left\|\left[I - \tilde{p}\left(\tilde{p}^H \tilde{p}\right)^{-1} \tilde{p}^H\right]\tilde{r}\right\|^2; \quad (17)$$
$$\overset{QR \ Decomposition}{\Longrightarrow} \ E_r(\lambda) = \left\|\left[I - QQ^H\right]\tilde{r}\right\|^2.$$

## 8. Terrestrial Communication Results

In this section the simulations are presented which depict the performance (BER) of the terrestrial wireless system. For these simulations, a rate 1/3 SCC has been employed, and formed by the unpunctured recursive convolutional component codes with the polynomial matrices for outer and inner codes. The re-ordering of the outer code's output is performed based on the algorithm presented in [30] which shown to provide additional coding gains for the SCCs. In this work, the interleaver size is 4096. For the simulations, the wireless communication system employs 16-ary QAM modulation with gray coded bit mapping, and it is assumed that the additive noise at the receiver is AWGN. **Figure 2** presents the performance for the discussed wireless system of interest with two transmitter and one receiver antennas (*i.e.* diversity order $D_O = 2 \times 1$

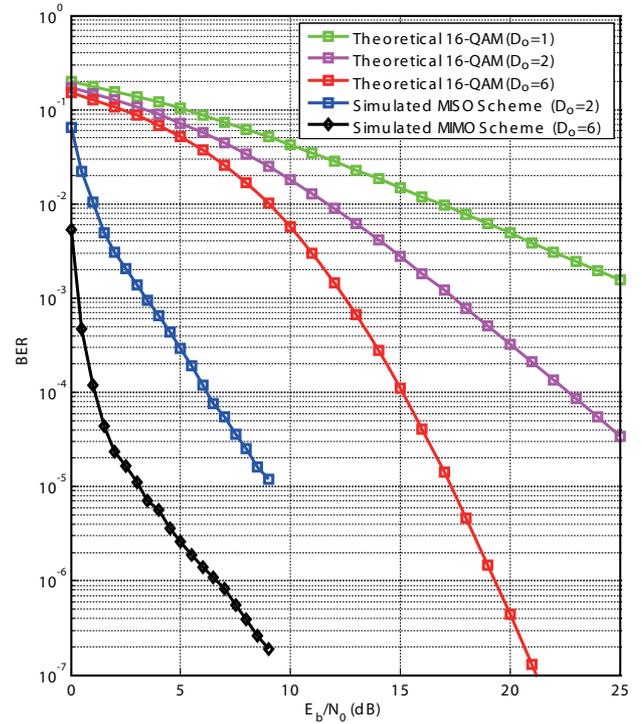

**Figure 2. Comparison of BER vs. ($E_b/N_0$) for coded 16-ary QAM MISO (2-Tx, 1-Rx) and MIMO (3-Tx, 2-Rx) terrestrial communication systems.**

= 2); and the system results have been compared with the BER curves associated with the theoretical system performance of the uncoded 16-QAM modulation in the Rayleigh multipath channel with $D_O = 1$ (*i.e.* SISO without diversity), and $D_O = 2$. The figure presents a good terrestrial system reference for comparison since in both approaches, the channel is assumed to be highly faded with blocked LoS. Hence, the additional coding of more than 15 dB implies that the employed technique yields astonishing additional coding gains at the expense of the tolerable system decoding complexities. Meanwhile, it should also be noted that the slope of the BER curve could be simply increased by increasing the component code's constraint length, the number of decoding iterations, and the interleaver size, at the expense of the delays which might not be acceptable for the hard real-time systems. In addition one more antenna at the transmitter and receiver has been employed in order to evaluate the coding gains due to the deployment of an additional antenna, while relaxing the other system simulation parameters for the designed SCC. In this system Tarokh scheme STBCs have been utilized ($D_O = 3 \times 2 = 6$), and compared with theoretical uncoded 16-QAM system over the Rayleigh faded channel with $D_O = 6$. As it can be seen from the figure, while the former scheme results in the BER of $10^{-5}$ at SNR = 9 dB, the latter yields the BER of $10^{-7}$, which should be considered as an immense terrestrial system advancement. Meanwhile, the introduced enhancement strategy to the wireless system is confirmed from the scatter diagrams (*i.e.* **Figure 3**) of the received constellation shapes which have shown that although the received symbols are highly faded for low the SNRs, they land at almost the same locations as in the transmitted constellations (these symbols are transmitted on $\pm 1 + \pm j$, $\pm 1 + \pm 3j$, $\pm 3 + \pm 3j$ and $\pm 3j + \pm 1j$). **Figures 4** and **5** present cross-correlation with sharp peak at the correct terrestrial time-delay with the time-lag corresponding to the minimum error as the actual time-delay, and the proposed relation between the SNR and the estimated time-delay in terrestrial environment, respectively.

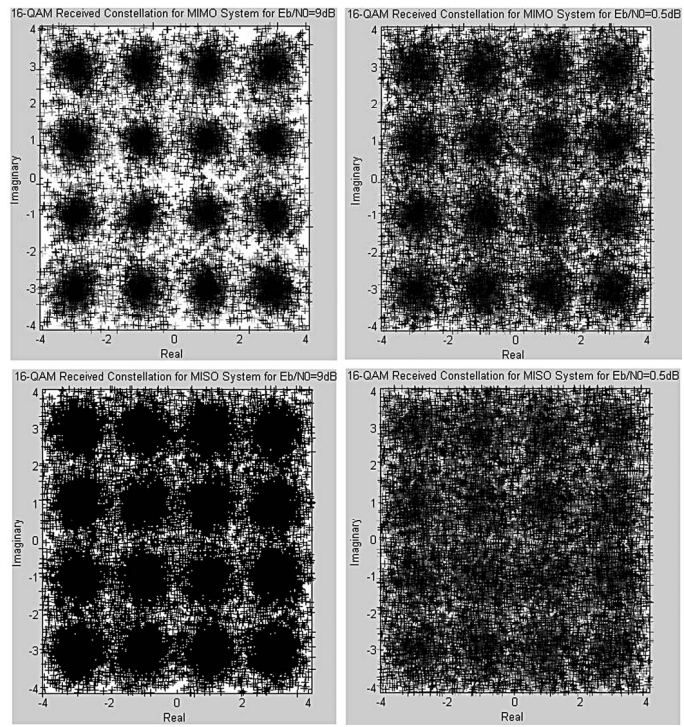

**Figure 3. Comparison of constellations at the MISO and MIMO wireless terrestrial communications receivers for M = 16-ary QAM digital system modulation scheme with gray coded bit assignment.**

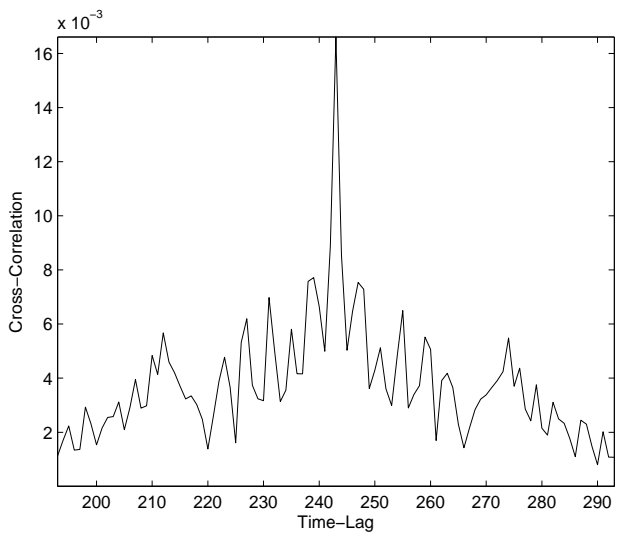

**Figure 4. Cross-correlation based on MLE in a simulated multipath terrestrial communications environment in presence of AWGN.**

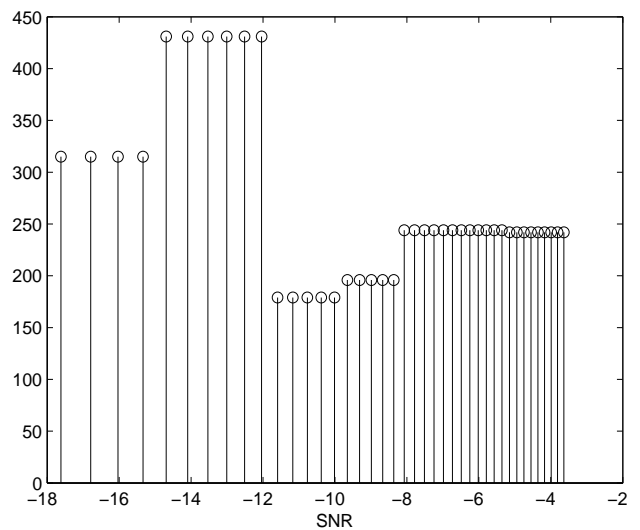

**Figure 5. The relation between SNR and estimated time-delays using MLE in a simulated multipath terrestrial communications system environment in presence of AWGN.**

## 9. Concluding Remarks

In this contribution, the performances of space-time codes based on Alamouti and Tarokh STBCs have been thoroughly evaluated in conjunction with high-performance turbo coding system schemes which were previously developed in order to yield astonishing system coding gains.

The introduced technique in this research could also be considered as a potential promising candidate for futuristic terrestrial systems and applications; wherein the channel environment is highly faded, and power and bandwidth are at premium so very low and high-modulation orders may not be adopted for the common highly-faded conditions of the terrestrial propagation environments.

As for the future research, efforts will be put into considering the proposed technique combined with the TCM for improving the system's efficiency. In this contribution, a promising mathematically-oriented analytical evaluation for the terrestrial system time-delay estimation (TDE) in presence of AWGN in the multipath environment has also been carried out. The presented analytical model can further be extended in order to analyze the terrestrial wireless communication system overall performance for the unknown number of radio paths in specified environment.